\documentclass[sigconf]{acmart}
\AtBeginDocument{%
  }

\usepackage[normalem]{ulem}
\usepackage{booktabs} 
\usepackage{subcaption}

\usepackage[english]{babel}
\usepackage[utf8x]{inputenc}
\usepackage[T1]{fontenc}
\usepackage{comment} 

\usepackage{amsmath}
\usepackage{graphicx}
\usepackage[table]{xcolor}
\usepackage[colorinlistoftodos]{todonotes}
\usepackage{multirow}
\usepackage{array} 
\usepackage{mwe}
\usepackage{xcolor, soul}
\sethlcolor{red}
\usepackage{siunitx}

\renewcommand\footnotetextcopyrightpermission[1]{}
\acmConference{}{}{}

\begin{document}

\title{CVT Archives and Chemical Embedding Measures for Multi-Objective Quality Diversity in Molecular Design}



\author{Dominic Mashak}
\affiliation{%
 \institution{Southwestern University}
 \city{Georgetown}
 \state{Texas}
 \country{USA}}
\email{mashakd@southwestern.edu}

\author{Jacob Schrum}
\affiliation{%
 \institution{Southwestern University}
 \city{Georgetown}
 \state{Texas}
 \country{USA}}
\email{schrum2@southwestern.edu}

\begin{abstract}
Nonlinear optical (NLO) materials are essential for photonic technologies, yet discovering optimal NLO molecules requires balancing multiple competing objectives across vast chemical spaces. Previous work showed that Multi-Objective MAP-Elites (MOME) with grid-based archives discovers diverse, high-quality molecules for electro-optic applications. However, uniform grid partitioning wastes archive capacity on chemically infeasible regions while undersampling high-density areas. We apply MOME with Centroidal Voronoi Tessellation (CVT) archives whose cells are defined by learned embeddings from ChemBERTa-2 Multi-Task Regression reduced via UMAP, capturing chemical similarity beyond simple structural features. We investigate a four-objective NLO molecular design problem: maximizing the $\beta / \gamma$ hyperpolarizability ratio, constraining HOMO-LUMO gap and linear polarizability to target ranges, and minimizing energy per atom. Our results demonstrate that embedding-based measures in CVT archives yield significantly higher median global hypervolume and multi-objective quality diversity scores, while filling nearly all native archive niches.
\end{abstract}



\maketitle

\setlength{\tabcolsep}{3pt}

\section{Introduction}

Nonlinear optical (NLO) materials encompass photonic devices such as electro-optic (EO) modulators, optical switches, and frequency converters~\cite{minasian:elsevier2005, saleh:photonics1991}. An effective EO modulator exploits the Pockels effect~\cite{saleh:photonics1991}, a second-order NLO process proportional to the first hyperpolarizability ($\beta$): high $\beta$ enables stronger modulation at smaller device footprints. However, practical performance requires optimizing four properties: a high $\beta/\gamma$ ratio to favor second-order over competing third-order optical responses (defined by $\gamma$); moderate linear polarizability ($\alpha \in [100,500]$~a.u.) for strong charge transfer without excessive optical loss or dispersion; a HOMO-LUMO gap ($\Delta E$) of $2$--$4$~eV, providing visible-range transparency while sustaining NLO-active charge-transfer character; and thermodynamic stability, proxied by minimizing total energy per heavy atom.

Previous work~\cite{mashak:gecco2026} systematically compares 
NSGA-II~\cite{deb:tec2002}, MAP-Elites~\cite{mouret:arxiv2015}, MOME~\cite{pierrot:gecco2022}, $(\mu{+}\lambda)$ evolution, and simulated annealing on this four-objective NLO problem using SMILES-encoded organic molecules evaluated with \textit{ab initio} Hartree-Fock (HF) calculations via PySCF~\cite{sun:jcphys2020} and the 3-21G basis set. 
MOME (Multi-Objective MAP Elites \cite{pierrot:gecco2022}) combines multiobjective and quality diversity (QD) optimization, 
seeking diverse genotypes exhibiting various objective tradeoffs.
Results showed that
MOME with a fine-grained uniform grid archive best populates diverse, high-quality niches, but exposed a key limitation: fixed grid cells waste archive capacity on chemically infeasible measure combinations (e.g., bond counts that cannot correspond to valid molecules) while undersampling high-density regions of chemical space~\cite{mashak:gecco2026}.

This work utilizes CVT-MOME, which replaces the fixed grid with a Centroidal Voronoi Tessellation (CVT) archive~\cite{vassiliades:tec2018} whose cells are defined by learned chemical embeddings. SMILES strings are encoded with ChemBERTa-2 MTR~\cite{ahmad:arxiv2022}, a transformer pre-trained on over 10 million PubChem compounds, then projected to a compact 10-dimensional manifold via UMAP~\cite{mcinnes:arxiv2018}. CVT centroids are seeded from this 
manifold, placing niches where molecules actually cluster in chemical space rather than in chemically uninhabited corners of a uniform grid. The result is an archive that partitions diversity along axes of genuine chemical similarity, providing more semantically coherent niches for evolutionary search, resulting in improved objective performance, as evidenced by the significantly higher global hypervolume score.

\section{Methods}


Molecule properties are calculated using the PySCF library\cite{sun:jcphys2020} and the same HF quantum chemistry method with 3-21G basis set used in previous work~\cite{mashak:arxiv2025}. SMILES (Simplified Molecular Input Line Entry System~\cite{weininger:jcics1988}) encodes molecular
structures as ASCII strings for evolution, and are also manipulated as in prior work~\cite{mashak:gecco2024, mashak:match2024}. We restrict molecules to C, N, O, and H atoms using single (\texttt{-}) and double (\texttt{=}) bonds only.
Canonical SMILES from RDKit~\cite{landrum:rdkit2010} ensure unique representations.
Mutation operations convert the parent SMILES to an editable molecular graph, apply the modification, and regenerate a canonical SMILES. Invalid results are discarded and retried up to 20 times before attempting a different mutation. Seven mutation operators are used: changing bond type, inserting an atom, adding a branch, deleting an atom, changing an atom type, adding a ring, and deleting a ring bond~\cite{mashak:gecco2026}. Crossover is not used to maintain chemical validity without complex repair mechanisms.

Multi-dimensional Archive of Phenotypic Elites (MAP-Elites)~\cite{mouret:arxiv2015} maintains an archive of elite solutions across a discretized measure space. Multiobjective MAP-Elites (MOME)~\cite{pierrot:gecco2022} extends MAP-Elites by storing local Pareto fronts within each archive bin rather than single elites. Each bin maintains a mutually non-dominated Pareto front of objective trade-offs. 
Solutions are generated by randomly sampling the archive, and assigned to bins based on their measures. They remain in their assigned bin if it is empty, or if they are non-dominated with respect to the previous occupants. Old occupants are only discarded if new occupants dominate them.

Grid-based MOME uses atom and bond counts as behavior descriptors.
Centroidal Voronoi Tessellation (CVT) archives~\cite{vassiliades:tec2018} replace the
uniform grid with $N$ Voronoi cells whose centroids $\{\mathbf{c}_k\}_{k=1}^{N}$ are
computed by $k$-means clustering~\cite{lloyd:tit1982} of a set of sample points in measure
space. Each molecule is assigned to the cell of its nearest centroid:
\begin{equation}
    k^* = \arg\min_k \| \mathbf{m} - \mathbf{c}_k \|_2
\end{equation}
where $\mathbf{m}$ is the molecule's measure vector. A KD-tree supports efficient
nearest-centroid lookup. Each cell stores a local Pareto front following the MOME
semantics.


The specific measure space we use with
CVT-MOME is based on learned embeddings capturing emergent chemical similarity.
For embedding, we use ChemBERTa-2 Multi-Task Regression (MTR)~\cite{ahmad:arxiv2022}, a BERT-style~\cite{devlin:naacl2019} transformer with 77 million parameters pre-trained on over 10 million SMILES strings from PubChem and fine-tuned on multiple molecular property regression tasks, giving its representations physicochemical grounding beyond the base language model.
A SMILES string is tokenized using a chemical-aware byte-pair encoding (BPE) tokenizer, then passed through 12 transformer layers producing contextual token embeddings of dimension 768. A single fixed-length molecular fingerprint $\mathbf{h}$ is obtained by mean pooling the final hidden states across all non-padding token positions:
\begin{equation}
    \mathbf{h} = \frac{\sum_{t=1}^{T} m_t \, \mathbf{h}_t}{\sum_{t=1}^{T} m_t}
\end{equation}
where $\mathbf{h}_t \in \mathbb{R}^{768}$ is the hidden state at position $t$, $m_t \in \{0,1\}$ is the attention mask, and $T$ is the sequence length. This yields $\mathbf{h} \in \mathbb{R}^{768}$ per molecule.


Because CVT archives require low-dimensional measures for efficient Voronoi lookup, the 768-dimensional fingerprints are reduced to $d=10$ dimensions using UMAP (Uniform Manifold Approximation and Projection~\cite{mcinnes:arxiv2018}) with cosine
distance ($n\_neighbors=30$, $min\_dist=0.1$). UMAP is fitted once on 10,000
randomly generated molecules at the start of each run, establishing a fixed manifold
for all subsequent embeddings.
The fitted embeddings of the initialization sample also seed CVT centroid
generation, ensuring centroids are placed in inhabited regions rather than empty ones.



\section{Experiment}

Code used to run our experiments is available at this url: \\ https://github.com/DominicMashak/Molecular-Evolution. These experiments focus on comparing MOME and CVT-MOME approaches to
evolving molecules for their NLO properties, and compare to NSGA-II~\cite{deb:tec2002}, a multiobjective but non-QD method.

MOME and CVT-MOME seed archives with 50 random molecules, then mutate randomly selected archive members for 1,800 iterations. NSGA-II uses $\mu = \lambda = 20$ over 90 generations. All experiments use 20 random seeds. Initial populations are generated from common scaffolds (e.g., C, C=C, C-C-N) by applying multiple random mutations to create valid, unique molecules. Molecules are restricted to 5-30 heavy atoms (non-hydrogen), forming a single connected component. All algorithms optimize four objectives: (1)~$\beta/\gamma$: maximize to favor second-order NLO responses over third-order; (2)~$f_{\alpha}$: deviation from target $\alpha \in [100,500]$~a.u.~\cite{piela:elsevier2020}; (3)~$f_{\Delta E}$: deviation from target HOMO-LUMO gap $2$--$4$~eV; and (4)~$E_{total}/N_{atoms}$: minimize total HF energy per heavy atom as a thermodynamic stability proxy; molecules with positive values (convergence failures or strained geometries) are rejected (Section~\ref{sec:medianglobalhv}).



Standard grid-based archives use two measures: $m_{1}$, the heavy-atom count (range 5--30), and $m_{2}$, the heavy-atom bond count where each bond is counted once regardless of type (range 4--32). Both measures are discretized evenly into a $20 \times 20$ grid (400 cells total).
This approach does not provide a one-to-one mapping between measure values and bins. 
Some adjacent measure values map to the same bin. 
Molecules with measure values beyond the boundaries are assigned to the nearest bin.


CVT-MOME uses $N=100$ centroids in the 10-dimensional UMAP embedding space. 
While the grid archive contains 400 potential cells, previous studies~\cite{mashak:gecco2026} indicate that fewer than 20\% of those cells are ever occupied due to the sparse distribution of valid chemical structures. We selected $N=100$ based on preliminary testing, which suggested that this centroid count provides a comparable number of "active" niches to the grid-based baseline.
Centroid generation is data-driven: the 10,000
UMAP-embedded molecules used to fit the manifold are directly passed to $k$-means as the sample set, so centroids are placed where molecules actually live in the manifold rather than in uninhabited regions. As a result, no archive capacity is wasted on chemically impossible combinations as the grid archive does (e.g., more bonds than atoms).

\section{Results}

MOME, CVT-MOME, and NSGA-II are compared in terms of the following results produced by each of their 20 distinct runs using different random seeds.
In general, results across runs do not follow a normal distribution, so we favor median scores to compare algorithm performance and assess statistical differences with the Kruskal-Wallis and Mann-Whitney U tests.

\subsection{Median Global Hypervolume}
\label{sec:medianglobalhv}

\begin{figure}[t]
\centering
\includegraphics[width=0.7\linewidth]{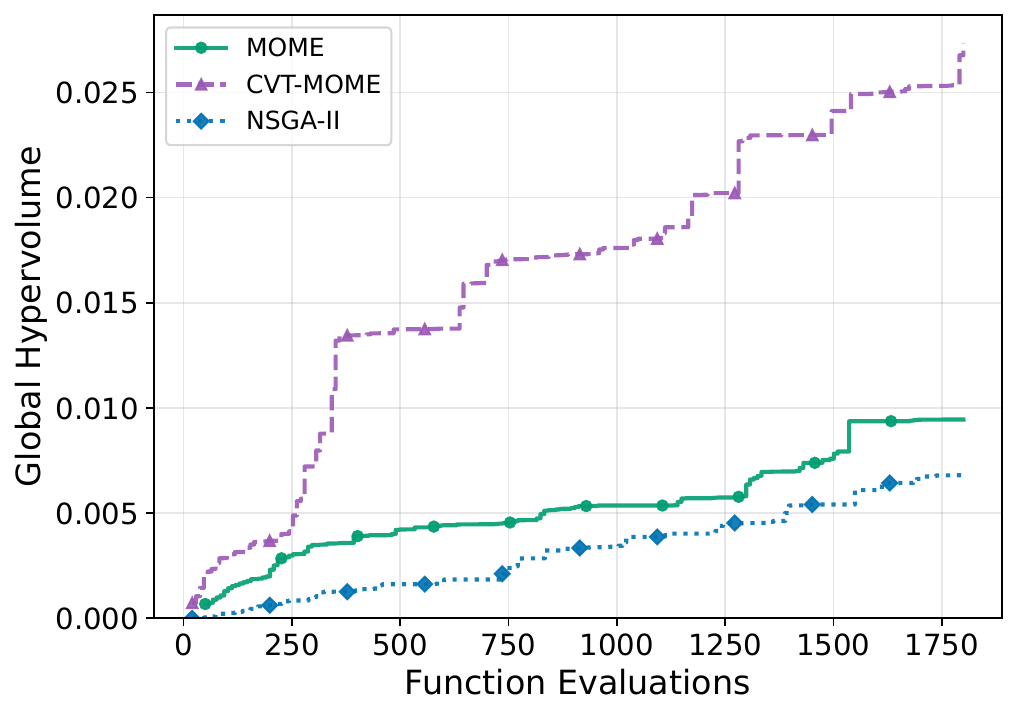}
\caption{\small Median global hypervolume across function evaluations (20 runs each).
CVT-MOME consistently achieves the highest median hypervolume throughout evolution.}
\Description[Median Global Hypervolume Scores Across Evaluations]{Median global hypervolume scores for MOME and CVT-MOME plotted against function evaluations across 20 runs.}
\label{fig:global_hv_eval}
\end{figure}

\begin{figure}[t]
\centering
\includegraphics[width=0.7\linewidth]{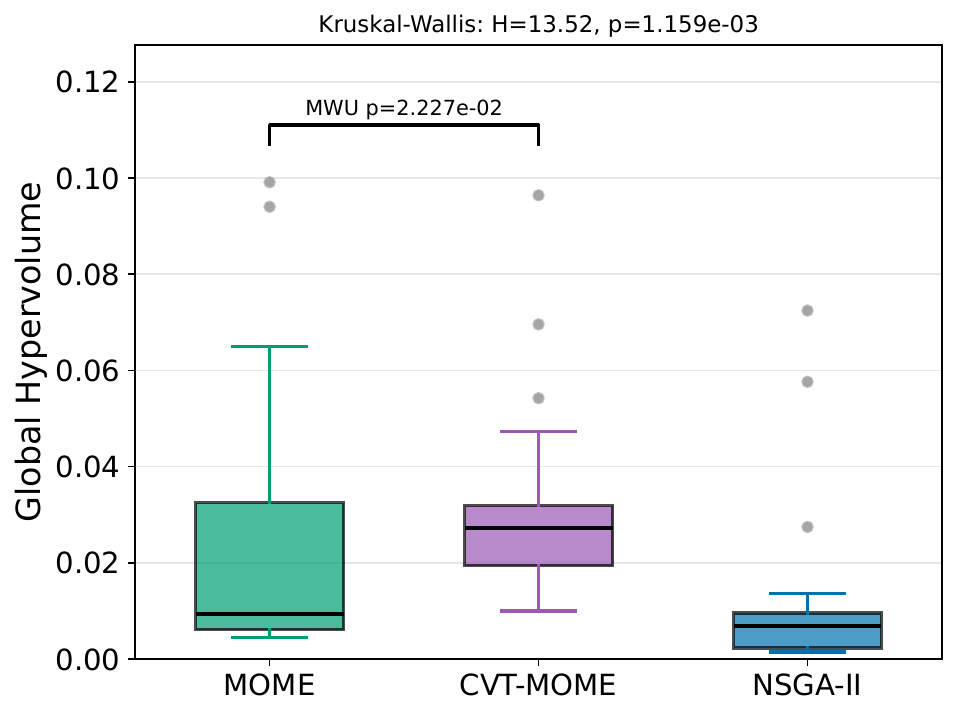}
\caption{\small Box-and-whisker plots of final global hypervolume across 20 runs per algorithm.}
\label{fig:global_hv_final}
\end{figure}

We compute a Pareto front across all molecules generated in each run and calculate its hypervolume (HV) using \texttt{pymoo}~\cite{blank:ieee2020}.
HV is the Lebesgue measure of the objective space dominated by the Pareto front relative to a fixed reference point~\cite{zitzler:tec2003}, providing a scalar summary of multiobjective quality.

However, hypervolume is sensitive to 
differences in objective scale and to extreme outliers \cite{zitzler:tec2003},
making normalization of objective scores necessary.
Furthermore, the HF calculations we use can sometimes be distorted by systemic errors \cite{szabo:1996}.

To ensure physical coherence, we prune any molecule whose \(\beta\) or \(\gamma\) violates the Kuzyk limit~\cite{kuzyk:jstqe2001}. For every candidate, we extract the total number of electrons \(N_e\) and the formal charge directly from its canonical SMILES string. The computed $\Delta E$ (converted to Hartree) serves as the first excitation energy \(E_{10}\). The bounds in atomic units are:
\begin{equation}
\beta_{\max} = 3^{1/4} \frac{N_e^{3/2}}{E_{10}^{7/2}}, \qquad
\gamma_{\max} = 4 \frac{N_e^{2}}{E_{10}^{5}}, \qquad
\gamma_{\min} = -0.25\,\gamma_{\max}
\end{equation}
Any molecule satisfying \(|\beta| > \beta_{\max}\) or \(\gamma \notin [\gamma_{\min},\gamma_{\max}]\) is discarded.

Some molecules also have extreme values for $\alpha$, HOMO-LUMO gap, and average energy,
and these extreme values would make a molecule unsuitable as an EO modulator despite
high $\beta/\gamma$.
Based on this knowledge, we discard molecules that exceed any of these specific property ranges: $\beta \in [0, 50000]$ and $\gamma \in [10, 10000]$~\cite{kanis:cr1994}. The lower bound of 10 on $\gamma$ is imposed to prevent numerical instability in the $\beta/\gamma$ objective: when $\gamma$ approaches zero, the ratio diverges, producing artificially large scores disconnected from genuine second-order NLO performance rather than reflecting a high $\beta$. These bounds are consistent with the plausible range of hyperpolarizability values for synthesizable organic chromophores~\cite{kanis:cr1994, mashak:arxiv2025}. We then normalize objective scores to the following ranges: $\beta/\gamma$: $[0, 500]$ (unitless), $f_{\alpha}$: $[0, 100]$ (a.u.), $f_{\Delta E}$: $[0, 8]$ (eV), $E_{total}/N_{atoms}$: $[-55, -10]$ (Hartree).
The $\beta/\gamma$ normalization ceiling of 500 caps scores at an achievable saturation point within the imposed absolute maximum of $50{,}000/10 = 5{,}000$, consistent with observations in viable organic NLO chromophores~\cite{kanis:cr1994}; the $f_{\alpha}$ bound of 100~a.u.\ caps the penalty at the target window half-width; $f_{\Delta E}$ of 8~eV spans C/N/O molecules under HF/3-21G~\cite{szabo:1996}; and the energy range $[-55, -10]$~Hartree reflects the per-heavy-atom HF/3-21G energetics of our candidate molecules, whose constituent atomic energies establish the physical lower bound while the upper bound excludes convergence failures~\cite{szabo:1996}. These ranges map each objective to $[0,1]$ before performing the HV calculation with reference point $\vec{0}$.

Figure \ref{fig:global_hv_eval} shows median hypervolume scores across evaluations. We call these \emph{global} hypervolumes to distinguish them from scores associated with MOQD (Section~\ref{sec:moqd}). CVT-MOME's median global hypervolume rises more steeply and reaches a substantially higher plateau than MOME's. By the end of evolution, CVT-MOME achieves a median normalized HV of $0.0273$ compared to $0.0095$ for MOME and $0.0068$ for NSGA-II. Figure~\ref{fig:global_hv_final} confirms this advantage holds across the full distribution of runs: a Kruskal-Wallis test yields $H = 13.52$, $p = 1.16 \times 10^{-3}$, and a Mann-Whitney U test between MOME and CVT-MOME gives $p = 2.23 \times 10^{-2}$, confirming statistically significant differences. CVT-MOME also exhibits lower inter-run variance than MOME, whose wide interquartile range reflects greater sensitivity to random seed.

\subsection{Median Multiobjective QD Score}
\label{sec:moqd}

\begin{figure*}[t]
\centering
\begin{subfigure}[t]{0.24\textwidth}
    \centering
    \includegraphics[width=\linewidth]{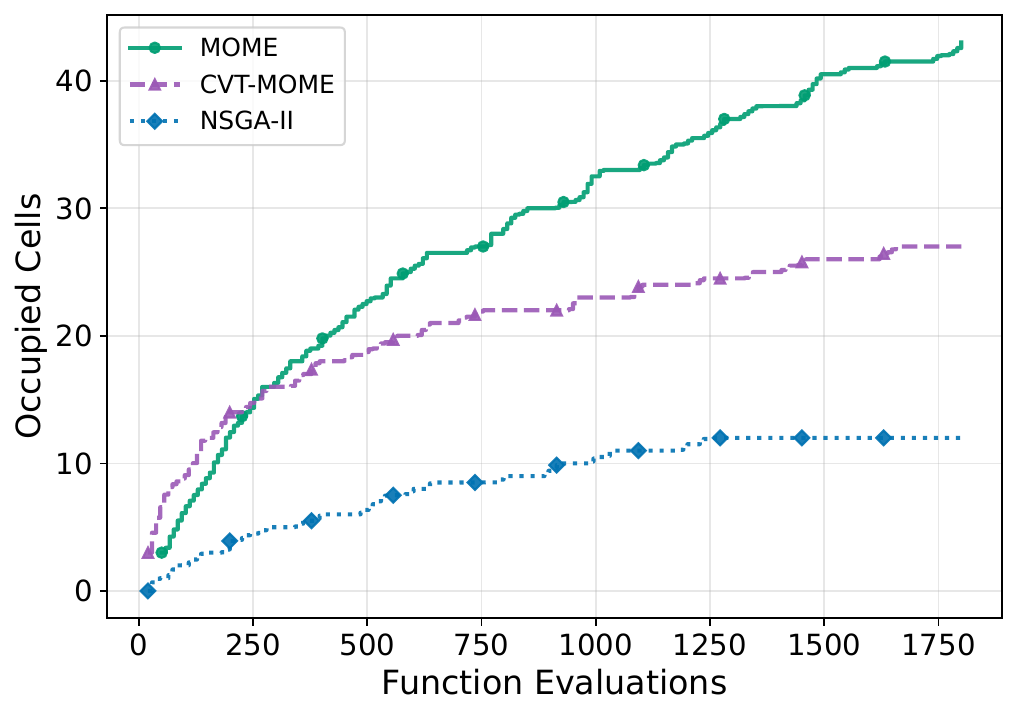}
    \caption{\small $\text{Count}_{Grid}$}
    \label{fig:count_f}
\end{subfigure}\hfill
\begin{subfigure}[t]{0.24\textwidth}
    \centering
    \includegraphics[width=\linewidth]{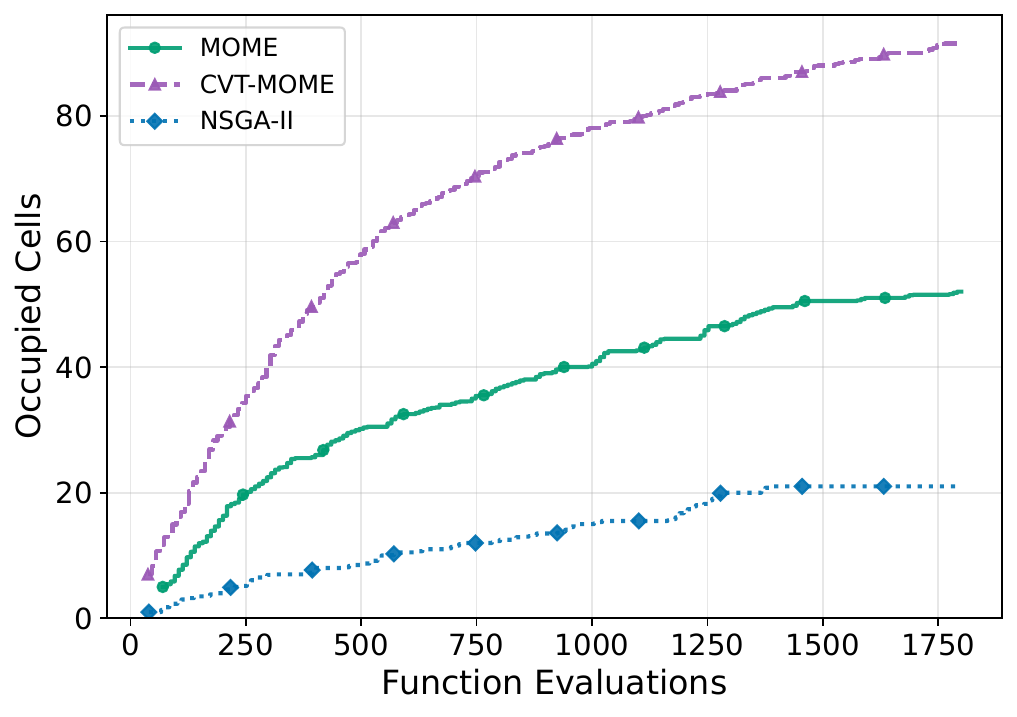}
    \caption{\small $\text{Count}_{CVT}$}
    \label{fig:count_c}
\end{subfigure}\hfill
\begin{subfigure}[t]{0.24\textwidth}
    \centering
    \includegraphics[width=\linewidth]{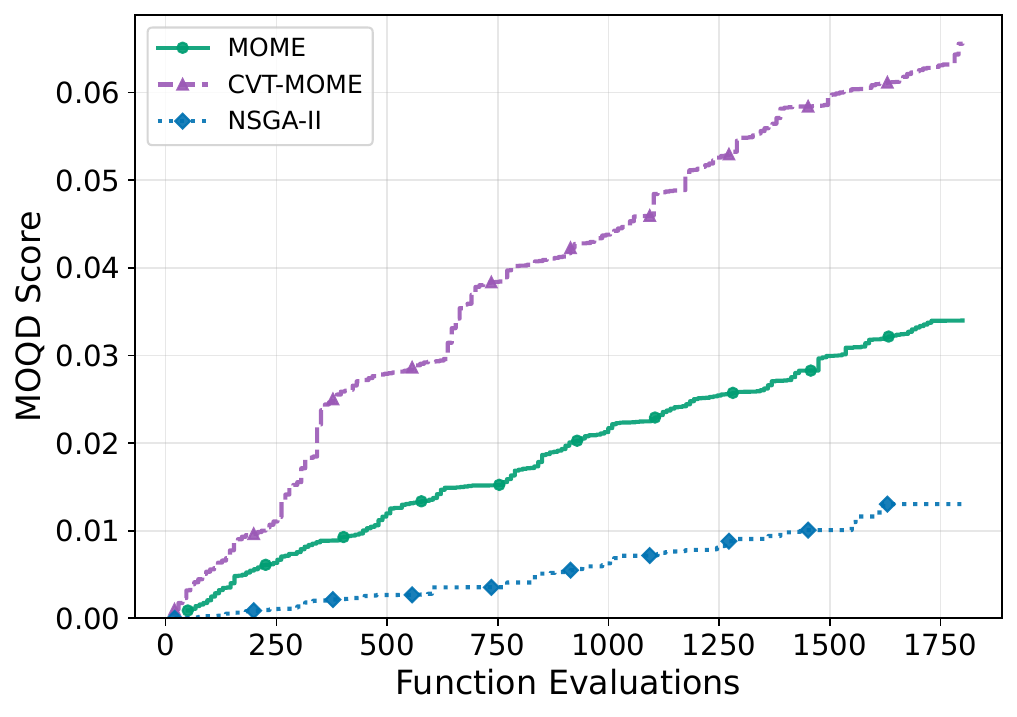}
    \caption{\small $\text{MOQD}_{Grid}$}
    \label{fig:moqd_f}
\end{subfigure}\hfill
\begin{subfigure}[t]{0.24\textwidth}
    \centering
    \includegraphics[width=\linewidth]{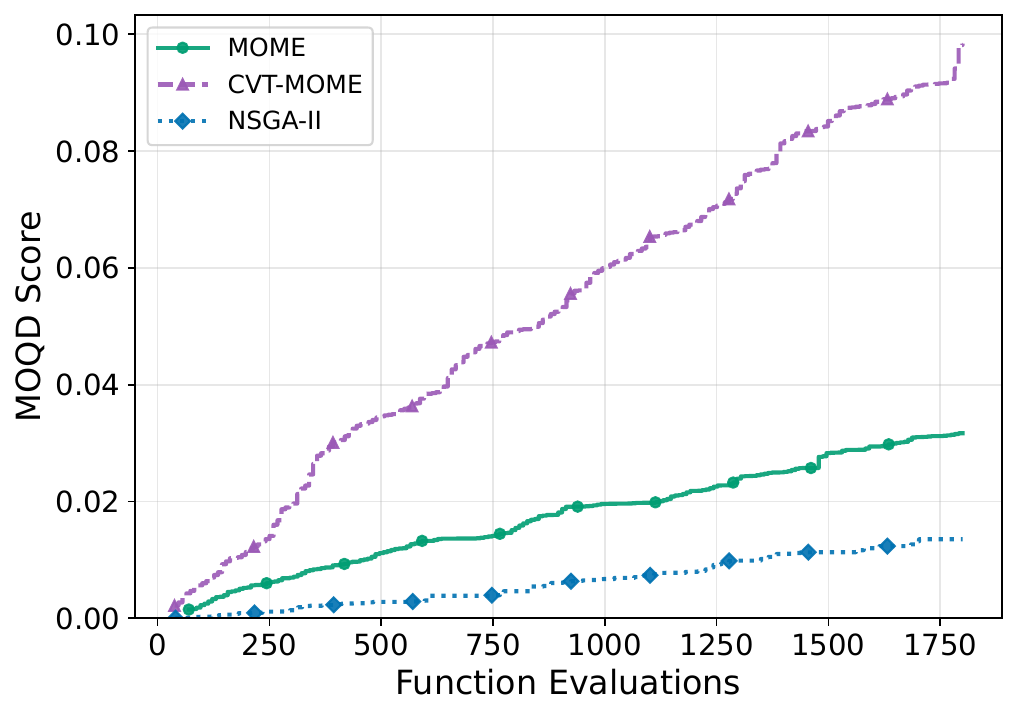}
    \caption{\small $\text{MOQD}_{CVT}$}
    \label{fig:moqd_c}
\end{subfigure}

\caption{\small Grid-based and CVT Archive Median Scores Across 20 Runs of Each Algorithm:
(\subref{fig:count_f}) Median bin count with grid-based archive. 
(\subref{fig:count_c}) Median bin count with CVT archive.
(\subref{fig:moqd_f}) Median MOQD with grid-based archive.
(\subref{fig:moqd_c}) Median MOQD with CVT archive.
}
\label{fig:fine_coarse_count_moqd}
\Description[Fine-grained and Coarse Archive Median Scores Across 20 Runs of Each Algorithm]{TODO}
\end{figure*}

\begin{figure}[t]
\centering
\begin{subfigure}[t]{0.15\textwidth}
    \centering
    \includegraphics[width=\linewidth]{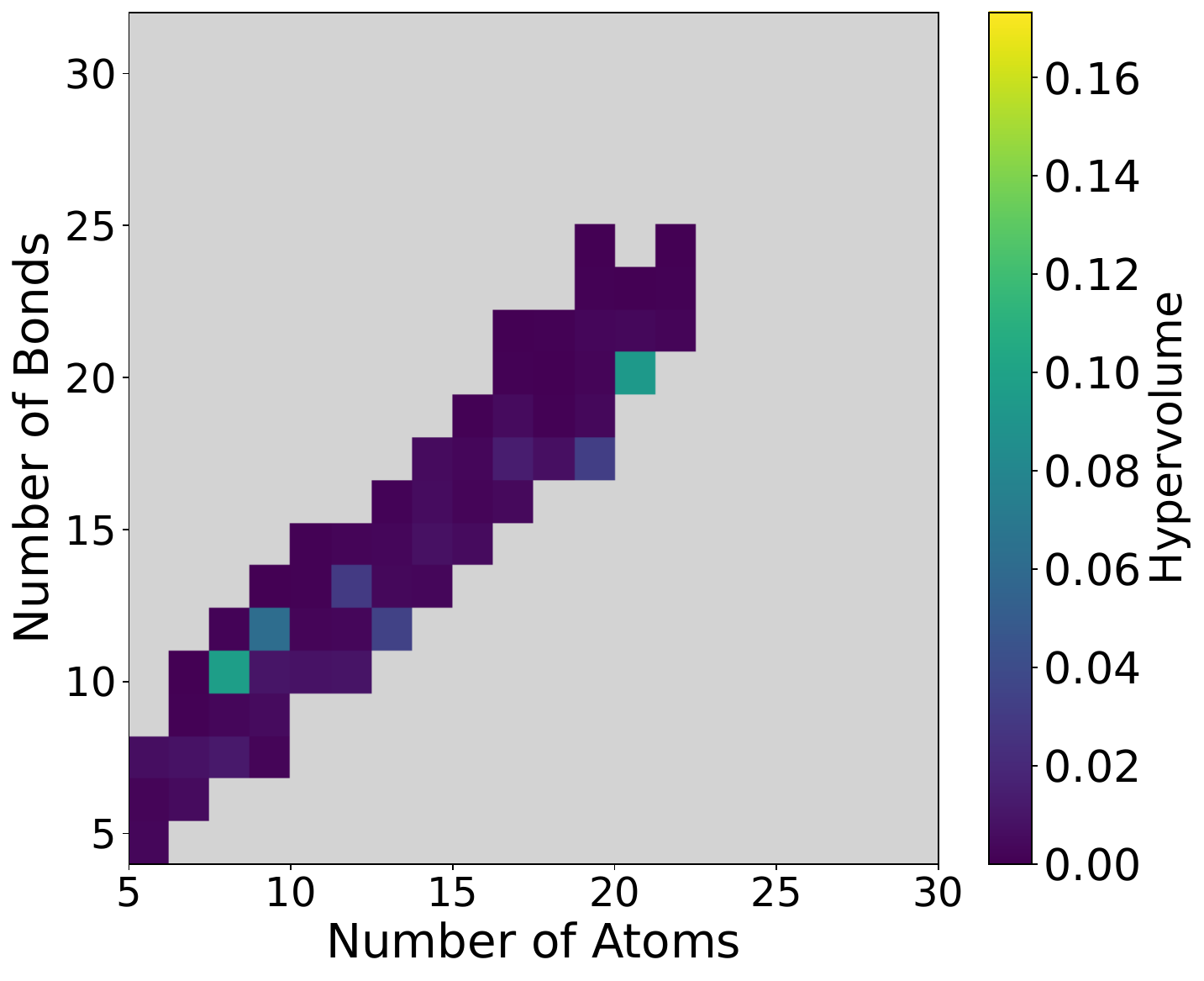}
    \caption{\small MOME}
    \label{fig:f_ml_hv_heat_mome}
\end{subfigure}
\hfill
\begin{subfigure}[t]{0.15\textwidth}
    \centering
    \includegraphics[width=\linewidth]{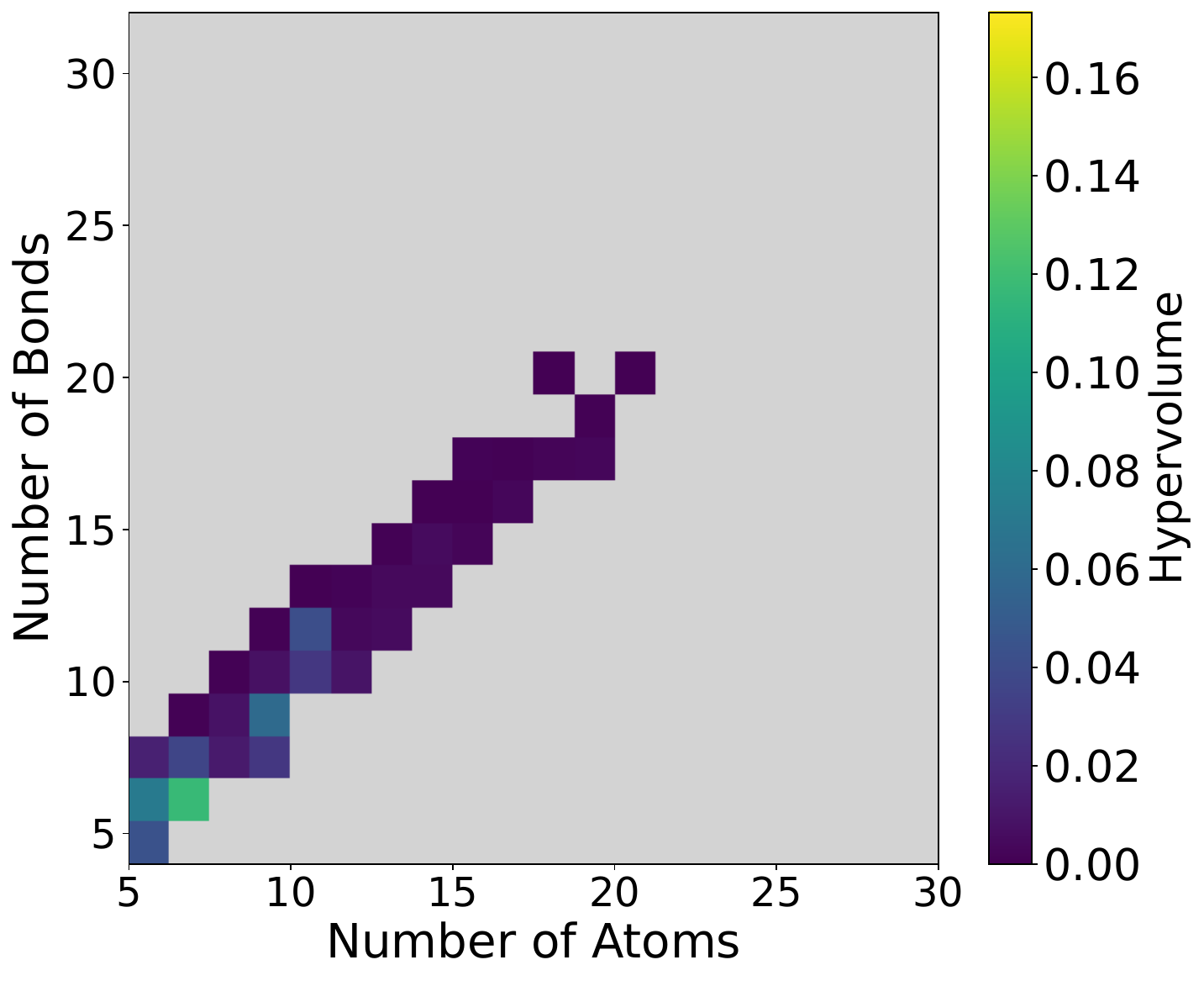}
    \caption{\small CVT-MOME}
    \label{fig:f_ml_hv_heat_cvt}
\end{subfigure}
\hfill
\begin{subfigure}[t]{0.15\textwidth}
    \centering
    \includegraphics[width=\linewidth]{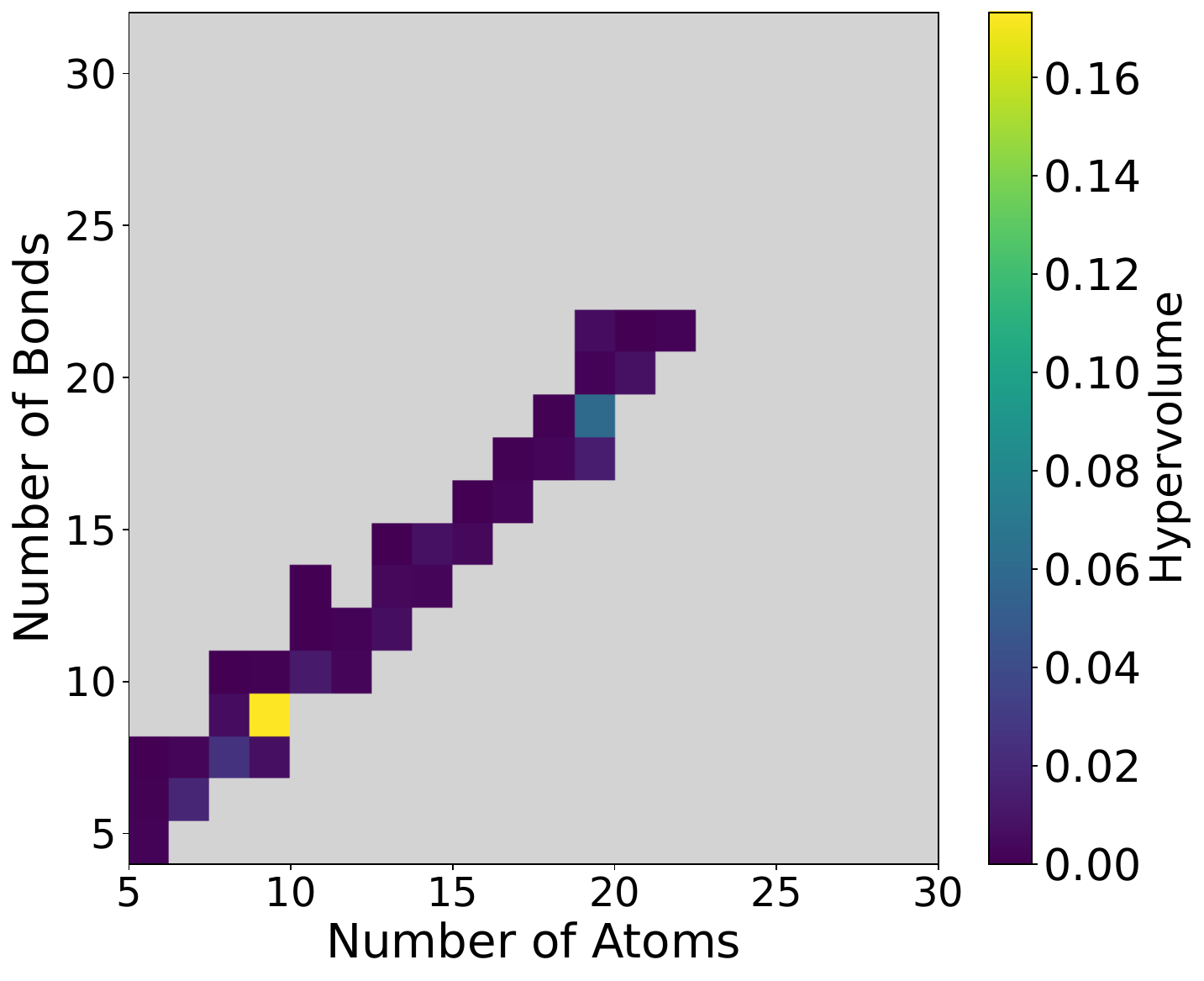}
    \caption{\small NSGA-II}
    \label{fig:f_ml_hv_heat_nsga2}
\end{subfigure}

\caption{\small Grid-based Mega Archive Hypervolume Heatmaps:
Archives pooling solutions from all 20 seeds per algorithm, with color scale showing each bin's HV score. The x-axis denotes the atom count, and the y-axis denotes the bond count.}
\label{fig:hv_heatmaps_f}
\Description[TODO]{TODO}
\end{figure}

The paper introducing MOME also introduced the MOQD score, or multiobjective QD score \cite{pierrot:gecco2022}. Recall that
MOME bins can contain multiple solutions representing a Pareto front across objectives. Because HV 
represents the quality of a Pareto front, the quality of a MOME bin is the
hypervolume of that bin, and MOQD score is the sum of all hypervolumes across all occupied bins
in the archive.

Figure~\ref{fig:count_f} shows that MOME consistently occupies more cells of the $20{\times}20$ grid archive than CVT-MOME, as its diversity pressure directly operates in atom-count $\times$ bond-count space. CVT-MOME maintains diversity in the 10-dimensional embedding space; when rebinned into the 2D grid, its solutions cluster in fewer structural bins. Rebinned into the CVT archive (Figure~\ref{fig:count_c}), CVT-MOME fills 91 of 100 centroids versus only 52 for MOME and 21 for NSGA-II, demonstrating broader chemical diversity that grid coverage alone fails to capture.

Despite occupying fewer grid cells, CVT-MOME achieves a substantially higher median MOQD score in the grid archive (Figure~\ref{fig:moqd_f}): $0.065$ for CVT-MOME versus $0.034$ for MOME and $0.013$ for NSGA-II. The cells CVT-MOME occupies contain Pareto fronts of considerably higher quality, more than compensating for reduced grid coverage. When MOQD is computed in the CVT archive (Figure~\ref{fig:moqd_c}), CVT-MOME's advantage grows further: $0.098$ versus $0.032$ for MOME and $0.014$ for NSGA-II.

\subsection{Mega Archive Hypervolume Heatmaps}

Figure~\ref{fig:hv_heatmaps_f} shows the mega-archive HV heatmaps, pooling all molecules from all 20 seeds. MOME distributes solutions broadly along the structural diagonal. CVT-MOME covers a somewhat narrower structural range but concentrates its highest-quality Pareto fronts at the small-molecule end, consistent with embedding-guided search pressure toward chemically promising regions. Notably, NSGA-II achieves the highest single-cell hypervolume among algorithms, reflecting its focused population pressure, though at the cost of the diversity that QD methods provide.

\section{Conclusions}

We demonstrate that CVT archives defined in a learned chemical embedding space substantially improve multi-objective quality diversity in NLO molecular design. CVT-MOME achieves a significantly higher median global hypervolume, grid-based MOQD, and CVT-based MOQD. While MOME occupies more cells in the structural grid archive, CVT-MOME fills almost all of its native CVT centroids compared to half for MOME, demonstrating substantially broader chemical diversity in embedding space. By placing niches where molecules actually cluster, using ChemBERTa-2 MTR embeddings projected to a 10-dimensional UMAP manifold, CVT-MOME avoids wasting archive capacity on structurally infeasible atom-count/bond-count combinations. These results demonstrate that embedding-informed archive structures significantly enhance the quality and diversity of discovered NLO molecules. Future work will explore this approach on drug-discovery tasks and compare it with other leading molecular optimization strategies.

\begin{acks}
The authors acknowledge that Generative AI (Claude) was used to refine the \texttt{matplotlib} code that produced all the result figures. 

\end{acks}

\bibliographystyle{ACM-Reference-Format}
\bibliography{mol-evo}

@inproceedings{mashak:gecco2024,
author = {Mashak, Dominic and Alexander, Steven},
title = {Finding Molecules with Specific Properties: Simulated Annealing vs. Evolution},
year = {2025},
isbn = {9798400714641},
nopublisher = {Association for Computing Machinery},
noaddress = {New York, NY, USA},
publisher = {ACM},
address = {NY},
nourl = {https://doi.org/10.1145/3712255.3726635},
nodoi = {10.1145/3712255.3726635},
longbooktitle = {Proceedings of the Genetic and Evolutionary Computation Conference Companion},
booktitle = {Genetic and Evolutionary Computation Conference Companion},
pages = {759–762},
numpages = {4},
nolocation = {NH Malaga Hotel, Malaga, Spain},
noseries = {GECCO '25 Companion}
}

@inproceedings{mashak:match2024,
author = {Mashak, Dominic and Alexander, Steven},
title = {Finding Molecules with Large
Hyperpolarizabilities},
year = {2025},
nopublisher = {MATCH Commun. Math. Comput. Chem.},
nourl = {https://match.pmf.kg.ac.rs/issues/m94n3/m94n3_25824.html},
nodoi = {10.46793/match94-3.25824},
booktitle = {MATCH Commun. Math. Comput. Chem.},
volume = {94},
noissue = {3},
nopages = {633-644},
}

@misc{mashak:arxiv2025,
      title={Benchmarking Hartree-Fock and DFT for Molecular Hyperpolarizability: Implications for Evolutionary Design}, 
      author={Dominic Mashak and S. A. Alexander},
      year={2025},
      eprint={2511.17767},
      archivePrefix={arXiv},
      primaryClass={physics.chem-ph},
      url={https://arxiv.org/abs/2511.17767}, 
}

@inproceedings{pierrot:gecco2022,
author = {Pierrot, Thomas and Richard, Guillaume and Beguir, Karim and Cully, Antoine},
title = {Multi-objective quality diversity optimization},
year = {2022},
isbn = {9781450392372},
nopublisher = {Association for Computing Machinery},
noaddress = {New York, NY, USA},
publisher = {ACM},
address = {NY},
nourl = {https://doi.org/10.1145/3512290.3528823},
nodoi = {10.1145/3512290.3528823},
longbooktitle = {Proceedings of the Genetic and Evolutionary Computation Conference},
booktitle = {Genetic and Evolutionary Computation Conference},
pages = {139–147},
numpages = {9},
nolocation = {Boston, Massachusetts},
noseries = {GECCO '22}
}

@misc{mouret:arxiv2015,
      title={Illuminating search spaces by mapping elites}, 
      author={Jean-Baptiste Mouret and Jeff Clune},
      year={2015},
      eprint={1504.04909},
      archivePrefix={arXiv},
      primaryClass={cs.AI},
      url={https://arxiv.org/abs/1504.04909}, 
}

@ARTICLE{deb:tec2002,
  author={Deb, K. and Pratap, A. and Agarwal, S. and Meyarivan, T.},
  journal={IEEE Transactions on Evolutionary Computation}, 
  title={A fast and elitist multiobjective genetic algorithm: NSGA-II}, 
  year={2002},
  volume={6},
  number={2},
  pages={182-197},
  doi={10.1109/4235.996017}
}

@article{sun:jcphys2020,
    author = {Sun, Q. and Zhang, X. and Banerjee, S. and Bao, P. and Barbry, M. and Blunt, N. S. and Bogdanov, N. A. and Booth, G. H. and Chen, J. and Cui, Zhi-Hao and Eriksen, J. J. and Gao, Y. and Guo, S. and Hermann, J. and Hermes, M. R. and Koh, K. and Koval, P. and Lehtola, S. and Li, Z. and Liu, J. and Mardirossian, N. and McClain, J. D. and Motta, M. and Mussard, B. and Pham, H. Q. and Pulkin, A. and Purwanto, W. and Robinson, P. J. and Ronca, E. and Sayfutyarova, E. R. and Scheurer, M. and Schurkus, H. F. and Smith, J. E. T. and Sun, C. and Sun, Shi-Ning and Upadhyay, S. and Wagner, L. K. and Wang, X. and White, A. and Whitfield, J. D. and Williamson, M. J. and Wouters, S. and Yang, J. and Yu, J. M. and Zhu, T. and Berkelbach, T. C. and Sharma, S. and Sokolov, A. Y. and Chan, G. Kin-Lic},
    title = {Recent developments in the PySCF program package},
    journal = {Journal of Chemical Physics},
    volume = {153},
    number = {2},
    nopages = {024109},
    year = {2020},
    nomonth = {07},
    issn = {0021-9606},
    nodoi = {10.1063/5.0006074},
    nourl = {https://doi.org/10.1063/5.0006074},
    noeprint = {https://pubs.aip.org/aip/jcp/article-pdf/doi/10.1063/5.0006074/16722275/024109_1_online.pdf},
}

@software{landrum:rdkit2010,
  author = {Landrum, Greg},
  title  = {RDKit: Open-source cheminformatics},
  year   = {2010},
  url    = {https://www.rdkit.org}
}

@ARTICLE{blank:ieee2020,
    author={J. {Blank} and K. {Deb}},
    journal={IEEE Access},
    title={pymoo: Multi-Objective Optimization in Python},
    year={2020},
    volume={8},
    number={},
    pages={89497-89509},
}

@article{weininger:jcics1988,
author = {Weininger, David},
title = {SMILES, a chemical language and information system. 1. Introduction to methodology and encoding rules},
journal = {Journal of Chemical Information and Computer Sciences},
volume = {28},
number = {1},
pages = {31-36},
year = {1988},
nodoi = {10.1021/ci00057a005},

noURL = { 
    
        https://doi.org/10.1021/ci00057a005
    
    

},
noeprint = { 
    
        https://doi.org/10.1021/ci00057a005
    
    

}

}

@ARTICLE{zitzler:tec2003,
  author={Zitzler, E. and Thiele, L. and Laumanns, M. and Fonseca, C.M. and da Fonseca, V.G.},
  journal={IEEE Transactions on Evolutionary Computation}, 
  title={Performance assessment of multiobjective optimizers: an analysis and review}, 
  year={2003},
  volume={7},
  number={2},
  pages={117-132},
  nodoi={10.1109/TEVC.2003.810758}}

@article{kanis:cr1994,
author = {Kanis, David R. and Ratner, Mark A. and Marks, Tobin J.},
title = {Design and construction of molecular assemblies with large second-order optical nonlinearities. Quantum chemical aspects},
journal = {Chemical Reviews},
volume = {94},
number = {1},
pages = {195-242},
year = {1994},
doi = {10.1021/cr00025a007},

URL = { 
    
        https://doi.org/10.1021/cr00025a007
    
    

},
eprint = { 
    
        https://doi.org/10.1021/cr00025a007
    
    

}

}

@incollection{piela:elsevier2020,
title = {The Molecule Subject to Electric or Magnetic Fields},
noeditor = {Lucjan Piela},
booktitle = {Ideas of Quantum Chemistry (Third Edition)},
publisher = {Elsevier},
noedition = {Third Edition},
pages = {253-335},
year = {2020},
isbn = {978-0-444-64248-6},
nodoi = {https://doi.org/10.1016/B978-0-44-464248-6.00012-0},
nourl = {https://www.sciencedirect.com/science/article/pii/B9780444642486000120},
author = {Lucjan Piela}
}

@incollection{minasian:elsevier2005,
title = {Modulation and Demodulation of Optical Signals},
noeditor = {Bob D. Guenther and Duncan G. Steel},
booktitle = {Encyclopedia of Modern Optics},
publisher = {Elsevier},
address = {Oxford},
pages = {129-138},
year = {2005},
author = {R.A. Minasian},
isbn = {978-0-12-814982-9},
nourl = {https://www.sciencedirect.com/referencework/9780123693952/encyclopedia-of-modern-optics}
}

@inbook{saleh:photonics1991,
author = {Bahaa E. A. Saleh and Malvin Carl Teich},
publisher = {John Wiley \& Sons, Ltd},
isbn = {9780471213741},
title = {Electro-Optics},
address = {NY},
booktitle = {Fundamentals of Photonics},
chapter = {18},
pages = {696--736},
doi = {https://doi.org/10.1002/0471213748.ch18},
url = {https://onlinelibrary.wiley.com/doi/abs/10.1002/0471213748.ch18},
noeprint = {https://onlinelibrary.wiley.com/doi/pdf/10.1002/0471213748.ch18},
year = {1991}
}

@book{szabo:1996,
  title={Modern Quantum Chemistry: Introduction to Advanced Electronic Structure Theory},
  author={Attila Szab{\'o} and Neil S. Ostlund},
  year={1996},
  publisher = {Dover Publications},
  address = {NY},
  nourl={https://api.semanticscholar.org/CorpusID:94743139}
}

@ARTICLE{kuzyk:jstqe2001,
  author={Kuzyk, M.G.},
  journal={IEEE Journal of Selected Topics in Quantum Electronics}, 
  title={Quantum limits of the hyper-Rayleigh scattering susceptibility}, 
  year={2001},
  volume={7},
  number={5},
  pages={774-780},
  nodoi={10.1109/2944.979338}}

@article{ahmad:arxiv2022,
  author    = {Ahmad, Walid and Simon, Elana and Chithrananda, Seyone and Grand, Gabriel and Ramsundar, Bharath},
  title     = {{ChemBERTa-2}: Towards Chemical Foundation Models},
  journal   = {arXiv preprint arXiv:2209.01712},
  year      = {2022},
}

@inproceedings{devlin:naacl2019,
  author    = {Devlin, Jacob and Chang, Ming-Wei and Lee, Kenton and Toutanova, Kristina},
  title     = {{BERT}: Pre-training of Deep Bidirectional Transformers for Language Understanding},
  booktitle = {Proceedings of the 2019 Conference of the North {A}merican Chapter of the Association for Computational Linguistics: Human Language Technologies},
  pages     = {4171--4186},
  year      = {2019},
  publisher = {Association for Computational Linguistics},
}

@article{vassiliades:tec2018,
  author    = {Vassiliades, Vassilios and Chatzilygeroudis, Konstantinos and Mouret, Jean-Baptiste},
  title     = {Using Centroidal {Voronoi} Tessellations to Scale Up the Multidimensional Archive of Phenotypic Elites Algorithm},
  journal   = {{IEEE} Transactions on Evolutionary Computation},
  volume    = {22},
  number    = {4},
  pages     = {623--630},
  year      = {2018},
}

@article{mcinnes:arxiv2018,
  author    = {McInnes, Leland and Healy, John and Melville, James},
  title     = {{UMAP}: Uniform Manifold Approximation and Projection for Dimension Reduction},
  journal   = {arXiv preprint arXiv:1802.03426},
  year      = {2018},
}

@article{lloyd:tit1982,
  author    = {Stuart P. Lloyd},
  title     = {Least Squares Quantization in {PCM}},
  journal   = {{IEEE} Transactions on Information Theory},
  volume    = {28},
  number    = {2},
  pages     = {129--137},
  year      = {1982},
}

@misc{mashak:gecco2026,
      title={Multi-Objective Evolutionary Design of Molecules with Enhanced Nonlinear Optical Properties}, 
      author = {Dominic Mashak and Jacob Schrum and S. A. Alexander},
      year={2026},
      eprint={2602.16044},
      archivePrefix={arXiv},
      primaryClass={physics.comp-ph}, 
}

\end{document}